\documentclass[12pt]{cernart}
\usepackage{epsfig}
\usepackage{amssymb}
\usepackage{cite}
\usepackage{colordvi}

\def\lapproxeq{\lower .7ex\hbox{$\;\stackrel{\textstyle
<}{\sim}\;$}}
\def\gapproxeq{\lower .7ex\hbox{$\;\stackrel{\textstyle
>}{\sim}\;$}}
\newcommand{\GeV}{\mathrm{GeV}}

\begin {document}
\dimen\footins=\textheight

\begin{titlepage}
\docnum{CERN--PH--EP/2006--040}
\date{20 December 2006}

\vspace{1cm}

\begin{center}
{\LARGE {\bf Spin asymmetry $A_1^d$ \\and the spin-dependent structure
function $g_1^d$ \\of the deuteron   
         at low values of $x$ and $Q^2$
              }}
\vspace*{0.5cm}
\end{center}

\author{\large The COMPASS Collaboration}
\vspace{2cm}

\begin{abstract}
 We present a precise 
measurement of the deuteron longitudinal spin 
asymmetry $A_1^d$ and of the deuteron spin-dependent structure
function $g_1^d$ at $Q^2 < 1~(\GeV/c)^2$ and 
$4\cdot10^{-5} < x < 2.5\cdot10^{-2}$ based on the data collected 
by the 
COMPASS experiment at CERN during the years 2002 and 2003. The
statistical precision is tenfold better than that of the previous
measurement in this region. The measured 
$A_1^d$ and $g_1^d$ are found to be consistent with zero in the whole
range of $x$. 
\vfill
\submitted{(Submitted to Physics Letters B)}
\end{abstract}

\newpage
%
%
\begin{Authlist}
{\large  The COMPASS Collaboration}\\[\baselineskip]
%
%
E.S.~Ageev\Iref{protvino},
V.Yu.~Alexakhin\Iref{dubna},
Yu.~Alexandrov\Iref{moscowlpi},
G.D.~Alexeev\Iref{dubna},
A.~Amoroso\Iref{turin},
B.~Bade\l ek\Iref{warsaw},
F.~Balestra\Iref{turin},
J.~Ball\Iref{saclay},
G.~Baum\Iref{bielefeld},
Y.~Bedfer\Iref{saclay}
% ...
\end{Authlist}
%
%
\Instfoot{bielefeld}{ Universit\"at Bielefeld, Fakult\"at f\"ur Physik, 33501 Bielefeld, Germany\Aref{f}}
\Instfoot{bochum}{ Universit\"at Bochum, Institut f\"ur Experimentalphysik, 44780 Bochum, Germany\Aref{f}}
\Instfoot{bonniskp}{ Universit\"at Bonn, Helmholtz-Institut f\"ur  Strahlen- und Kernphysik, 53115 Bonn, Germany\Aref{f}}
\Instfoot{bonnpi}{ Universit\"at Bonn, Physikalisches Institut, 53115 Bonn, Germany\Aref{f}}
\Instfoot{brno}{Institute of Scientific Instruments, AS CR, 61264 Brno, Czech Republic\Aref{g}}
\Instfoot{burdwan}{ Burdwan University, Burdwan 713104, India\Aref{i}}
\Instfoot{calcutta}{ Matrivani Institute of Experimental Research \& Education, Calcutta-700 030, India\Aref{j}}
\Instfoot{dubna}{ Joint Institute for Nuclear Research, 141980 Dubna, Moscow region, Russia}
\Instfoot{erlangen}{ Universit\"at Erlangen--N\"urnberg, Physikalisches Institut, 91054 Erlangen, Germany\Aref{f}}
\Instfoot{freiburg}{ Universit\"at Freiburg, Physikalisches Institut, 79104 Freiburg, Germany\Aref{f}}
\Instfoot{cern}{ CERN, 1211 Geneva 23, Switzerland}
\Instfoot{heidelberg}{ Universit\"at Heidelberg, Physikalisches Institut,  69120 Heidelberg, Germany\Aref{f}}
\Instfoot{helsinki}{ Helsinki University of Technology, Low Temperature Laboratory, 02015 HUT, Finland  and University of Helsinki, Helsinki Institute of  Physics, 00014 Helsinki, Finland}
\Instfoot{licerec}{Technical University in Liberec, 46117 Liberec, Czech Republic\Aref{g}}
\Instfoot{lisbon}{ LIP, 1000-149 Lisbon, Portugal\Aref{h}}
\Instfoot{mainz}{ Universit\"at Mainz, Institut f\"ur Kernphysik, 55099 Mainz, Germany\Aref{f}}
\Instfoot{miyazaki}{University of Miyazaki, Miyazaki 889-2192, Japan\Aref{k}}
\Instfoot{moscowlpi}{Lebedev Physical Institute, 119991 Moscow, Russia}
\Instfoot{munichlmu}{Ludwig-Maximilians-Universit\"at M\"unchen, Department f\"ur Physik, 80799 Munich, Germany\Aref{f}}
\Instfoot{munichtu}{Technische Universit\"at M\"unchen, Physik Department, 85748 Garching, Germany\Aref{f}}
\Instfoot{nagoya}{Nagoya University, 464 Nagoya, Japan\Aref{k}}
\Instfoot{praguecu}{Charles University, Faculty of Mathematics and Physics, 18000 Prague, Czech Republic\Aref{g}}
\Instfoot{praguectu}{Czech Technical University in Prague, 16636 Prague, Czech Republic\Aref{g}}
\Instfoot{protvino}{ State Research Center of the Russian Federation, Institute for High Energy Physics, 142281 Protvino, Russia}
\Instfoot{saclay}{ CEA DAPNIA/SPhN Saclay, 91191 Gif-sur-Yvette, France}
\Instfoot{telaviv}{ Tel Aviv University, School of Physics and Astronomy, 
              69978 Tel Aviv, Israel\Aref{l}}
\Instfoot{triestictp}{ ICTP--INFN MLab Laboratory, 34014 Trieste, Italy}
\Instfoot{triest}{ INFN Trieste and University of Trieste, Department of Physics, 34127 Trieste, Italy}
\Instfoot{turin}{ INFN Turin and University of Turin, Physics Department, 10125 Turin, Italy}
\Instfoot{warsaw}{ So{\l}tan Institute for Nuclear Studies and Warsaw University, 00-681 Warsaw, Poland\Aref{m} }
\Instfoot{warsawtu}{ Warsaw University of Technology, Institute of Radioelectronics, 00-665 Warsaw, Poland\Aref{n} }
\Anotfoot{a}{Also at IST, Universidade T\'ecnica de Lisboa, Lisbon, Portugal}
\Anotfoot{b}{Also at University of East Piedmont, 15100 Alessandria, Italy}
\Anotfoot{c}{On leave of absence from JINR Dubna}               
\Anotfoot{d}{Also at Chubu University, Kasugai, Aichi, 487-8501 Japan}
\Anotfoot{e}{Also at Yamagata University, Yamagata, 992-8510 Japan}
\Anotfoot{f}{Supported by the German Bundesministerium f\"ur Bildung und Forschung}
\Anotfoot{g}{Suppported by Czech Republic MEYS grants ME492 and LA242}
\Anotfoot{h}{Supported by the Portuguese FCT - Funda\c{c}\~ao para
               a Ci\^encia e Tecnologia grants POCTI/FNU/49501/2002 and POCTI/FNU/50192/2003}
\Anotfoot{i}{Supported by DST-FIST II grants, Govt. of India}
\Anotfoot{j}{Supported by  the Shailabala Biswas Education Trust}
\Anotfoot{k}{Supported by the Ministry of Education, Culture, Sports,
               Science and Technology, Japan; Daikou Foundation and Yamada Foundation}
\Anotfoot{l}{Supported by the Israel Science Foundation, founded by the Israel Academy of Sciences and Humanities}
\Anotfoot{m}{Supported by KBN grant nr 621/E-78/SPUB-M/CERN/P-03/DZ 298 2000,
               nr 621/E-78/SPB/CERN/P-03/DWM 576/2003-2006, and by MNII reasearch funds for 2005--2007}
\Anotfoot{n}{Supported by  KBN grant nr 134/E-365/SPUB-M/CERN/P-03/DZ299/2000}

\vfill

\hbox to 0pt {~}
%
%
\end{titlepage}


%

In the nucleon structure investigations by high energy lepton probes,
the region of low $x$ corresponds to  high parton densities,
where new dynamical mechanisms may be revealed. The longitudinal
structure function $g_1(x,Q^2)$ is presently the only observable
which permits the study of low $x$ processes in spin dependent
interactions. The existing data have been obtained
exclusively from fixed-target experiments where the low values of $x$
strongly correlate with low values of $Q^2$. Therefore
theoretical interpretations of the results require a suitable
extrapolation of the parton ansatz to the low-$Q^2$ region and possibly
also an inclusion of nonperturbative mechanisms, which vanish at higher $Q^2$.

Contrary to the spin-independent structure functions, the small-$x$
behaviour of both the singlet
and the non-singlet part of $g_1$ is controlled by double
logarithmic terms, {\it i.e.} by those terms which correspond to powers of
ln$^2$(1/$x$) at each order of the perturbative expansion \cite{ryskin}.
The double logarithmic effects go beyond the  DGLAP
evolution and can be accommodated in it using special techniques
\cite{techniques,jkbz,greco}.
Different approaches permit
 a smooth extrapolation of the obtained $g_1$ to the low-$Q^2$ region
\cite{jkbb_ln2,greco}
where it may also be complemented  by a non-perturbative component
\cite{nonpert}.
The double logarithmic terms generate the leading small-$x$ behaviour
of $g_1$ where the relevant Regge poles are expected to have a low
intercept.

The region of low $x$ and fixed $Q^2$ is the Regge limit
of the (deep) inelastic scattering where the Regge pole exchange model
should be applicable. In this model the shape of
$g_1$ at $x\rightarrow$ 0 ({\it i.e.} at $Q^2\ll W^2$ where $W^2$ is the
$\gamma^* N$ centre-of-mass energy squared) is parametrised as
\begin{equation}
g_1^i(x,Q^2) \sim \beta(Q^2)x^{-\alpha_i(0)}.
\label{regge}
\end{equation}
Here the index $i$ refers to the flavour singlet ($s$) and nonsinglet ($ns$)
combinations of proton and neutron structure functions and
$\alpha_i(0)$ denotes the Regge trajectory function at zero momentum
transfer.
It is expected that $\alpha_{s,ns}(0) \lapproxeq 0$ and that
$\alpha_s(0)\approx \alpha_{ns}(0)$ \cite{hei}. This behaviour
of $g_1$ should translate to the $W^{2\alpha}$ dependence of the
Compton cross-section at $Q^2\rightarrow 0$ where $g_1$ should be a finite
function of $W^2$, free from any kinematical singularities or zeros.

The spin-dependent structure function of the deuteron $g_1^d(x, Q^2)$
has been accurately measured in
the perturbative region, $Q^2 > 1~(\GeV/c)^2$
\cite{smc,e143,e155_d,hermes_new,
compass_a1_recent}. Due to the relatively low incident energy, the
deep inelastic scattering  events collected in {those} experiments
cover only a limited range of $x$.
The behaviour of $g_1$ at $x\lapproxeq 0.001$ in the large-$Q^2$ region
is unknown due to the lack of data from colliders with polarised beams.

\begin{figure}
\begin{minipage}{0.9\linewidth}
\begin{center}
\includegraphics[width=\textwidth,clip]{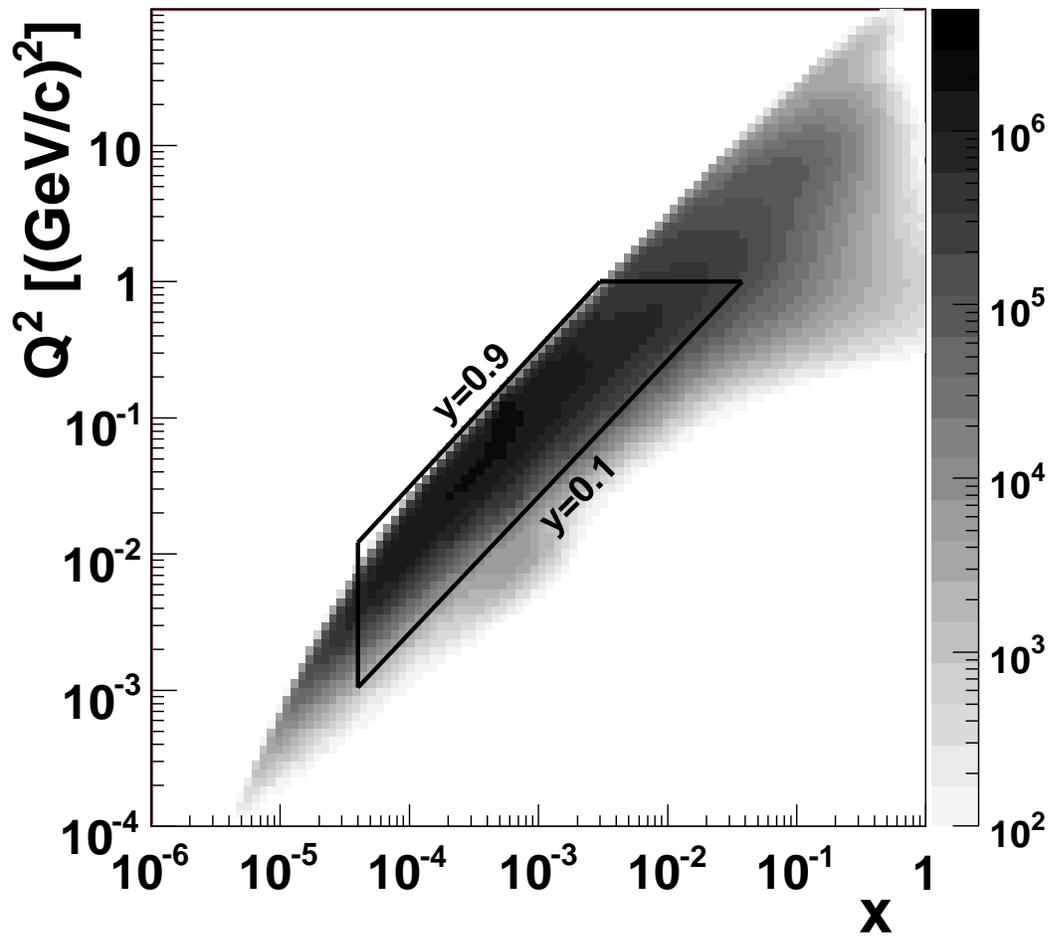}
\end{center}
\caption{\small COMPASS acceptance in the ($x$, $Q^2$)  plane. The contour
indicates the region selected for this analysis.} 
\label{fig:acc_new}
\end{minipage}
\end{figure}

Measurements  at low $x$ and low $Q^2$ put very high demands on event 
triggering and reconstruction and are very scarce: they
were performed only by the SMC at CERN on proton and deuteron
targets \cite{smc_lowx}.  
{Here} we present new results from the COMPASS experiment at CERN 
on the deuteron longitudinal spin
asymmetry $A_1^d$ and the spin-dependent structure function $g_1^d$ in the range
 $0.001<Q^2< 1~(\GeV/c)^2$ in the photon virtuality and 
$4\cdot 10^{-5}<x<2.5\cdot10^{-2}$ in
the Bjorken scaling variable. This range is essentially the same as 
that covered by the SMC \cite{smc_lowx}, but the present measurements 
result in about tenfold better precision. They complement
our recently published measurements   obtained 
in the region $0.004 < x < 0.7$ and $1< Q^2 < 100~(\GeV/c)^2$ \cite{compass_a1_recent}. 
The data were collected during the years 2002 and 2003.
They cover the kinematic range presented in Fig.~\ref{fig:acc_new}.
We refer the reader to reference \cite{mallot} for the description of  the 
160~GeV/$c$ positive muon beam, the two-cell $^6$LiD polarised target and the 
COMPASS spectrometer and to Ref.~\cite{ms_phd} for a detailed description
of the analysis.

The COMPASS data acquisition system is triggered by coincidence
signals in hodoscopes, defining the direction of the scattered
muon behind the spectrometer magnets and/or by a signal in the hadron 
calorimeters \cite{trigger}. Triggers due to halo muons are
suppressed by veto counters installed upstream of the target.
COMPASS uses three types of triggers: $i)$ inclusive ones, based on muon
detection only $ii)$ semi-inclusive triggers, based on muon detection and
presence of energy deposit in the hadron calorimeters and 
$iii)$ a calorimetric trigger where only information from
the hadron calorimeters is used. The low-$x$ and low-$Q^{2}$ region is dominated
by semi-inclusive triggers. The contribution of the inclusive ones 
is below 5\% for $x<0.001$ and exceeds 30\% only for $x > 0.01$. Also the 
contribution of the standalone calorimetric trigger  is negligible there. In
the kinematic region considered here events are characterised by small muon
scattering angles and their kinematics may be distorted by real photon 
emission.
Therefore in the analysis presented here the so-called hadron method \cite{smc} 
is used. This means that all events in our sample 
require the presence of the trajectories of a reconstructed
beam muon, a scattered muon and at least one additional outgoing particle, 
together defining an interaction point.
The presence of hadrons in the final state improves the
reconstruction of the interaction point and reduces the background
of events originating from radiative processes and from the muons scattered
off atomic electrons.
It has been checked that the use of the hadron method does not bias 
the inclusive asymmetries \cite{smc}. 

The momentum of the incoming muon, centred around 160~GeV/$c$ and
measured in the beam spectrometer, is required 
to be between 140 and 180~GeV/$c$. The reconstructed interaction point
has to be located inside one of the target cells. In addition,
the extrapolated beam muon trajectory is required to cross entirely both 
target cells in order to equalize the flux seen by  each of them.
 The scattered muon is identified by detectors situated behind hadron 
absorbers and its trajectory must be consistent with the
hodoscope signals used for the event trigger.

Events  are selected by cuts on the four-momentum transfer
squared, $Q^2 < 1~(\GeV/c)^2$, the fractional energy of the virtual photon,
$ 0.1 < y < 0.9$, and the scaling variable $x > 4\cdot10^{-5}$.
The remaining cuts are the same as those used in the  $Q^{2} > 1~(\GeV/c)^2$
analysis \cite{compass_a1_recent}, with additional
quality checks on the interaction point, appropriate to the present 
kinematics \cite{ms_phd}. According to the hadron method
we also require the most energetic hadron having  $z_h > 0.1$
($z_h$ is a fraction of the virtual photon energy in the
laboratory frame, carried by a hadron).

\begin{figure}
\begin{center}
\includegraphics[width=\textwidth,clip]{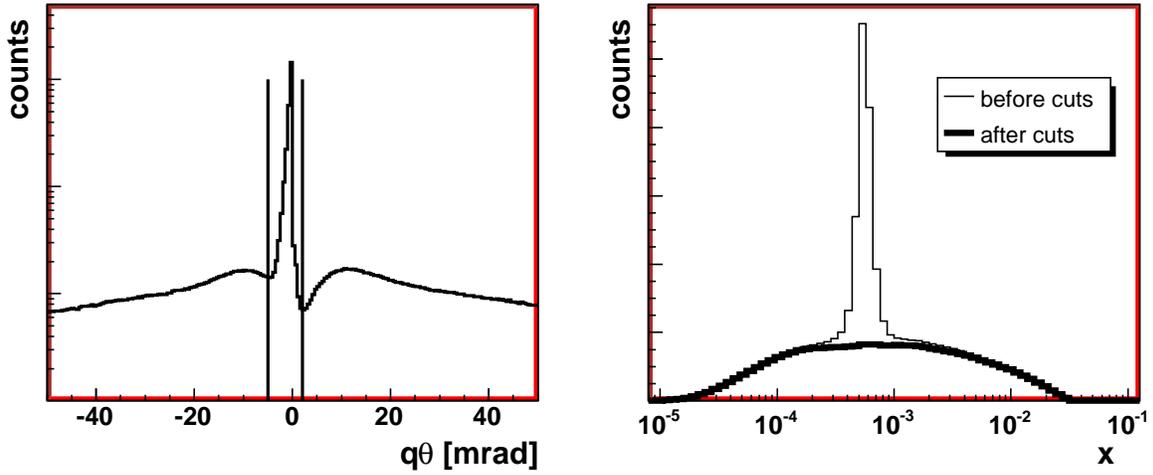}
\end{center}
\caption{\small Removal of the $\mu^+ e^- \rightarrow \mu^+ e^-$ 
scattering events. 
{\bf Left:} distribution of the variable $q\theta$ (see text for 
the definition) for events with one (positive or
negative) hadron candidate outgoing from the primary interaction point. 
Events between vertical lines are removed from further analysis. 
Note the logarithmic scale on the vertical axis.
{\bf Right:} $x$ distribution of events with one negative hadron candidate, 
before and after $\mu e$  event rejection.}
\label{fig:mue}
\end{figure}

At low values of $x$ the sample is contaminated by events
of muon elastic scattering off atomic electrons, 
$\mu^+ e^-\rightarrow \mu^+ e^-$,
occurring at $x_{{\mu}e} = m_{\mbox{\scriptsize electron}}/M = 5.45\cdot10^{-4}$ 
($M$ is the proton mass) 
and at very small scattering angles.
To remove such events, cuts are imposed on
a variable, $q\theta$, defined as the product of
the angle $\theta$ between the virtual photon and the hadron candidate
and the sign $q$ of the electric charge of the hadron.
Depending upon the number of hadron candidates outgoing from the
interaction point,
the event is rejected if $-5 < q\theta < 2$~mrad or  $-2 < q\theta < 0$~mrad 
depending whether it contains one or two hadron candidates \footnote{A part
of that condition, $0 < q\theta < 2$~mrad, encompasses misidentified
muons and beam halo muons.}. 
The distribution of the $q\theta$ variable and the $x$ spectrum
before and after the  $\mu e$ scattering  rejection are
presented in Fig.~\ref{fig:mue}. 
The background of  $\mu e$
events which remains under the elastic peak is estimated to be smaller
than 1\% of the data sample. 
As the electromagnetic calorimeter (ECAL) could not be fully used
in the present analysis, the cuts used for $\mu e$ scattering rejection 
presented above are applied in the whole range of $x$ to reduce the yield 
of unwanted radiative events.
A study using a small subsample of events where the ECAL was available shows
that around 50\% of those unwanted events are excluded from the data sample
in this way.  The remaining  background of radiative
events accounts for less than about 1\% of the data sample.

\begin{figure}
  \begin{minipage}{0.53\linewidth}
    \hskip-2cm
    \includegraphics[width=\textwidth,clip]{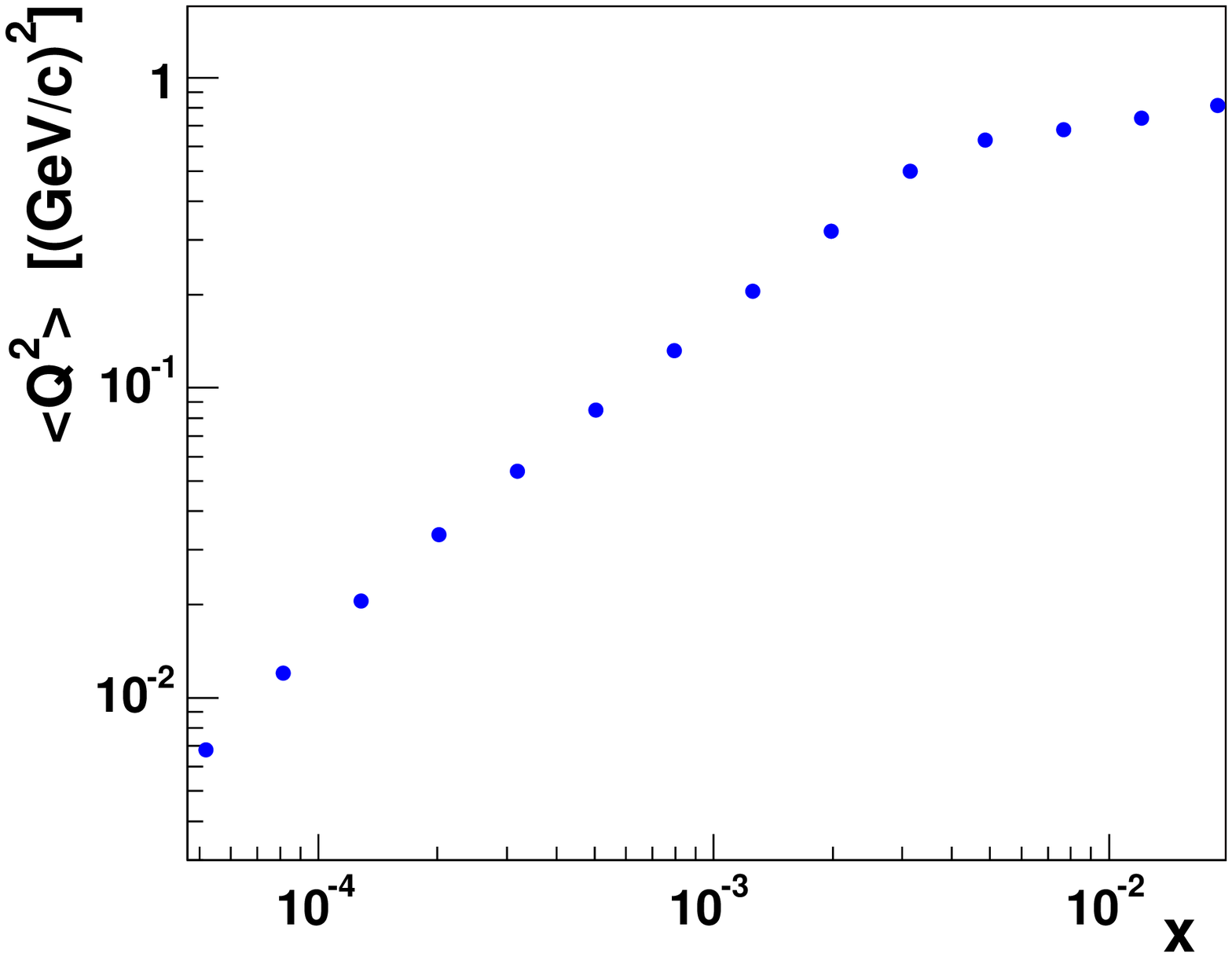}
    \caption{\small $\langle Q^2\rangle$ as a function of $x$
for the final data sample.}
    \label{fig:xq2_mean}
  \end{minipage}
  \hfill
  \begin{minipage}{0.52\linewidth}
    \hskip-1cm
    \includegraphics[width=\textwidth,clip]{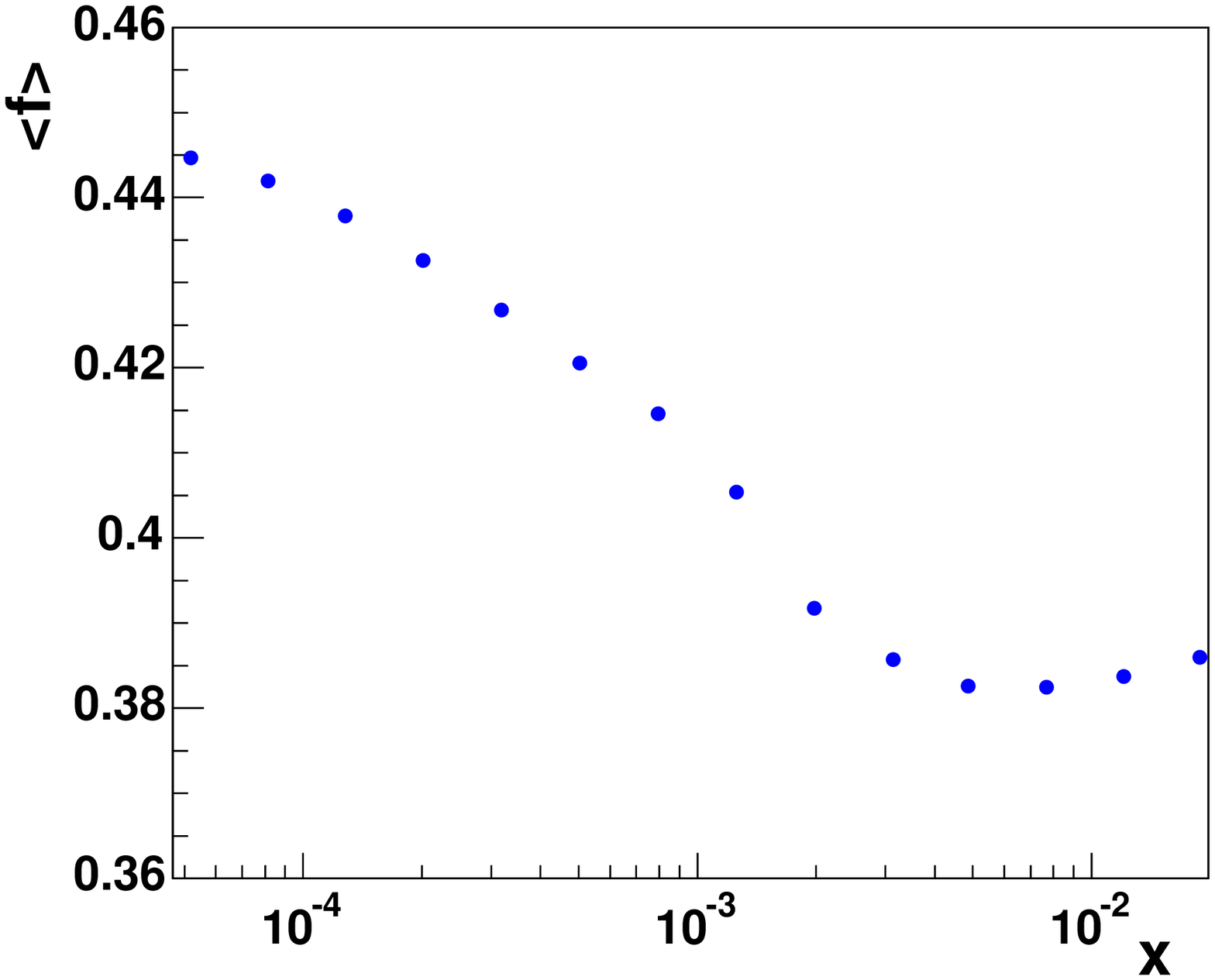}
    \caption{\small Mean effective dilution factor, $\langle f\rangle$, as 
a function of $x$ for the final data sample.}
    \label{fig:dilfac}
  \end{minipage}
\end{figure}

The resulting sample consists of 280 million events, out of which  
about 40\% were obtained in 2002. This is about 200 times more than
in Ref.~\cite{smc_lowx}.
The acceptance in the ($x$, $Q^2$) plane after all the cuts is
shown by the contour superimposed on Fig.~\ref{fig:acc_new}. Average values 
of $Q^2$ in bins of $x$ are presented in Fig.~\ref{fig:xq2_mean}.

During  data taking the two target cells are polarised in opposite directions, 
so that the deuteron spins are parallel or antiparallel
to the spins of the incoming muons. The spins are inverted
every 8 hours by a rotation of the target magnetic field. In 2002 and 2003 
the average beam and target deuteron polarisations were about $-0.76$  
and $\pm 0.51$, respectively.

Extraction of the cross-section asymmetry $A_1^d$
in the kinematic region where $Q^2$ extends down to about
0.001~$(\GeV/c)^2$ demands special care. The common practise of neglecting 
the $m_{\mu}^2/Q^2$ terms ( $m_{\mu}$ is the muon mass) 
in the expression for the cross-sections cannot be applied in this region. 
Therefore we present below the general spin formalism where all the 
$m_{\mu}^2/Q^2$ terms are properly taken into account. 
The only approximation applied is neglecting the 
$m_\mu^2/E^2$ terms ($E$ is the incident muon energy) 
which are of the order of $10^{-7}$ at our kinematics. 
In all the formulae we consider the exchange of one virtual photon only. 
The interference effects between virtual $Z^0$ and photon exchange in the 
inelastic muon scattering have been measured in an unpolarised
experiment \cite{bcdms} and found negligible in the
kinematic range of current fixed target experiments (see also \cite{smc_97}).


The polarised inelastic muon--deuteron inclusive
scattering cross-section $\sigma$ in the
one-photon exchange approximation can be written as the sum of a
spin-independent term $\bar{\sigma}$ and a
spin-dependent term $\Delta \sigma $ and involves the muon
helicity $h_{\mu}=\pm 1$
\begin{equation}
  \sigma = \bar{\sigma} - \frac{1}{2} h_{\mu} \Delta\sigma.
\end{equation}
Eight independent structure functions parametrise the cross-section
for a spin-1 target; this is twice as many as for the spin-1/2 case.
Apart of the spin-independent structure
functions $F_1$ and $F_2$ and the spin-dependent structure functions
$g_1$ and $g_2$,  four additional structure
functions,  $b_1$, $b_2$, $b_3$, $b_4$ 
are needed in the spin-1 case \cite{b14}.
All these functions depend on $Q^2$
and $x$. 
Following previous analyses, cf.\ Refs.~\cite{smc,smc_lowx,compass_a1_recent} 
we  neglect $b_{1-4}$ since they are predicted to be small \cite{b14}.
Then the expressions for the cross-sections
$\bar\sigma$ and $\Delta\sigma$ and thus the cross-section asymmetries
$A_\parallel$ and $A_\perp$ become identical to those for a spin-1/2 target. 

The spin-independent cross-section for parity-conserving interactions
can be expressed in terms of two unpolarized structure functions
$F_1$ and $F_2$:
\begin{equation}
\nonumber
\bar{\sigma}\equiv\frac{{\rm d^2}\bar{\sigma}} {{\rm d}x {\rm d}Q^2} =
\frac{4 {\pi}{\alpha}^2}{Q^{4} x}
\Biggl[ xy^2\left (1-\frac{2m_{\mu}^2}{Q^2}\right ) F_1(x,Q^2) + \\
\left (1 - y - \frac{{\gamma}^2 y^2}{4}\right ) F_2(x,Q^2) \Biggr],
\end{equation}
where 
\begin{equation}
\gamma = {2Mx \over \sqrt{Q^2}} = {\sqrt{Q^2} \over \nu}
\label{eq:gamma}
\end{equation}
and $\nu$ is the energy of the exchanged virtual photon.

When the muon spin and the deuteron spin form an angle $\psi$,
the cross-section $\Delta\sigma$ can be expressed as~\cite{jaf_g2}
\begin{equation}
 \Delta \sigma=\cos\psi\,\Delta \sigma_{\parallel}
  + \sin\psi\,\cos\phi\,\Delta \sigma_{\perp}.
\label{delta_sigma}
\end{equation}
Here $\phi$ is the azimuthal angle between the scattering plane
and the spin plane.
The cross-sections $\Delta \sigma_{\parallel}$ and $\Delta \sigma_{\perp}$
refer to the two configurations where the
deuteron spin is (anti)parallel or orthogonal to the muon spin;
$\Delta \sigma_{\parallel}$ is the difference
between the cross-sections for antiparallel and parallel spin orientations
and $\Delta \sigma_{\perp} = \Delta \sigma_{\rm T}/\cos\phi$,
the difference between the cross-sections at angles
$\phi$ and $\phi + \pi$.
The corresponding differential cross-sections, which can be written
in terms of the two structure functions $g_1$ and $g_2$, are given by

\begin{equation}
  \Delta\sigma_{\parallel}\equiv
  \frac{{\rm d}^2\Delta\sigma_{\parallel}}{{\rm d}x {\rm d}Q^2}
  = \frac{16\pi\alpha^2 y}{Q^4} \left[ \left (1 - {y\over 2}
  -{\gamma^2 y^2\over 4} - {{m^2_{\mu} y^2}\over{Q^2}}\right )g_1 -
{\gamma^2 y\over 2} g_2 \right]
\end{equation}
                                                                                       
\noindent
and
                                                                                       
\begin{equation}
\Delta\sigma_{\rm T} = \cos\phi\Delta\sigma_{\perp}\equiv
 \frac{{\rm d}^3\Delta\sigma_{\rm T}}{{\rm d}x {\rm d}Q^2{\rm d}\phi}
 =  \cos\phi\,\frac{8\alpha^2 y}{Q^4}\,\gamma\,
 \sqrt{1-y-{\gamma^2y^2 \over 4}}
 \left[ {y \over 2} \left (1 + {{2 m^2_{\mu}} \over {Q^2}} y\right ) 
g_1 + g_2 \right].
\end{equation}

The relevant asymmetries are

\begin{equation}
      A_{\parallel} = {\Delta\sigma_{\parallel} \over 2 \bar{\sigma}},
   \hspace{1cm}
             A_{\perp}     = {\Delta\sigma_{\perp} \over 2 \bar{\sigma}}.
\label{Asy_cross}
\end{equation}

The cross-section asymmetry $A^d_\parallel = (\sigma^{\uparrow \downarrow} - 
\sigma^{\uparrow \uparrow}) /
(\sigma^{\uparrow \downarrow} + \sigma^{\uparrow \uparrow})$, for antiparallel
($\uparrow \downarrow$) and parallel ($\uparrow \uparrow$)  spins of the 
incoming muon and the target deuteron can be obtained from the numbers of 
events $N_i$ 
collected from each cell before and after reversal of the target spins:
\begin{equation}
N_i = a_i \phi_i n_i {\overline \sigma} (1 + P_B P_T f A^d_\parallel),
~~~~~i=1,2,3,4,
\label{ni}
\end{equation}
where $a_i$ is the acceptance, $\phi_i$ the incoming flux, $n_i$ the number 
of target nucleons, 
$P_B$ and $P_T$ the beam and target polarisations and $f$ the 
effective target dilution factor.
The latter includes a corrective factor $\rho=\sigma_d^{1\gamma}/\sigma_d^{tot}$ 
\cite{terad} accounting for radiative events on the {unpolarised} deuteron
and a correction for the relative polarisation of deuterons bound in $^6$Li 
compared to free deuterons. Average values of $f$ in bins of
$x$ for the final data sample are presented in Fig.~\ref{fig:dilfac}. 
The extraction of the  
spin asymmetry was performed
as in Refs.~\cite{cmp23,compass_a1_recent}. The four relations of 
Eq.~(\ref{ni}), corresponding to the two cells and two spin orientations, lead
to a second-order equation in $A^d_\parallel$. This method ensures that
fluxes and acceptances cancel out in the asymmetry calculation on the condition 
that the ratio of the acceptances of the two cells is the same before and after spin 
reversal, cf.\ Ref.~\cite{smc_97}.

The longitudinal and transverse virtual-photon deuteron asymmetries, 
$A_1^d$ and $A_2^d$, are defined 
{via the asymmetry of absorption cross-sections of transversely polarised photon}
as
\begin{equation}
A_1^d = (\sigma_0^T - \sigma_2^T) / (2 \sigma^T),\hspace{1cm} 
A_2^d = (\sigma_0^{TL} + \sigma_1^{TL})/(2 \sigma^T),
\label{a12d}
\end{equation}
where $\sigma_J^T$ is the $\gamma^{*}$-deuteron absorption cross-section for a total
spin projection $J$ in the photon direction, $\sigma_J^{TL} $ 
results from the interference between transverse and longitudinal
amplitudes for $J = 0,1$ and 
$\sigma^T = (\sigma_0^T+ \sigma_1^T + \sigma_2^T)/3$ is the total transverse
photoabsorption cross-section.
The relation between $A_1^d, A_2^d$ and the experimentally measured 
$A^d_\parallel$, $A^d_\perp$  is 
\begin{equation}
A^d_\parallel = D(A^d_1 + \eta A^d_2) ,\hspace{1cm} A^d_\perp = d(A^d_2 - \xi A^d_1) ,
\label{ad}
\end{equation}
where $D$ (the so called depolarisation factor), $\eta$, $d$ and $\xi$ 
depend on kinematics:

\begin{equation}
D= {{ y\left[ (1 + {{\gamma^2y}/{2}}) (2 - y) -  {{2y^2 m_{\mu}^2} /
{Q^2}}  \right] }
\over{ y^2 (1- {{2m_{\mu}^2}/{Q^2}})(1+\gamma^2) + 2(1+R)
( 1-y - {{\gamma^2y^2} /{4}}) }} ,
\end{equation}

\begin{equation}
\eta =
{{\gamma \left( 1 - y - {{\gamma^2y^2}/{4}}  - { {{y^2 m_{\mu}^2}/{Q^2}}}  \right)}\over{(1 + {{\gamma^2y}/{2}})(1- {{y}/{2}}) - {{y^2 m_{\mu}^2}/{Q^2}}} } ,
\label{dep}
\end{equation}

\begin{equation}
d = {{\sqrt{ 1 - y - {{\gamma^2y^2}/{4}} }  ~(1+{{\gamma^2y}/{2}})}\over{ (1-{{y}/{2}})(1 +{ {\gamma^2 y}/{2}} ) - {{y^2 m_{\mu}^2}/{Q^2}}  }} D ,
\end{equation}

\begin{equation}
\xi =  {{\gamma  (1 - {{y}/{2}} -  {{y^2 m_{\mu}^2}/{Q^2}}) } \over{ 1 + { {\gamma^2 y}/{2}}} } .
\end{equation}

\begin{figure}
  \begin{minipage}{0.53\linewidth}
    \hskip-2cm
    \includegraphics[width=\textwidth,clip]{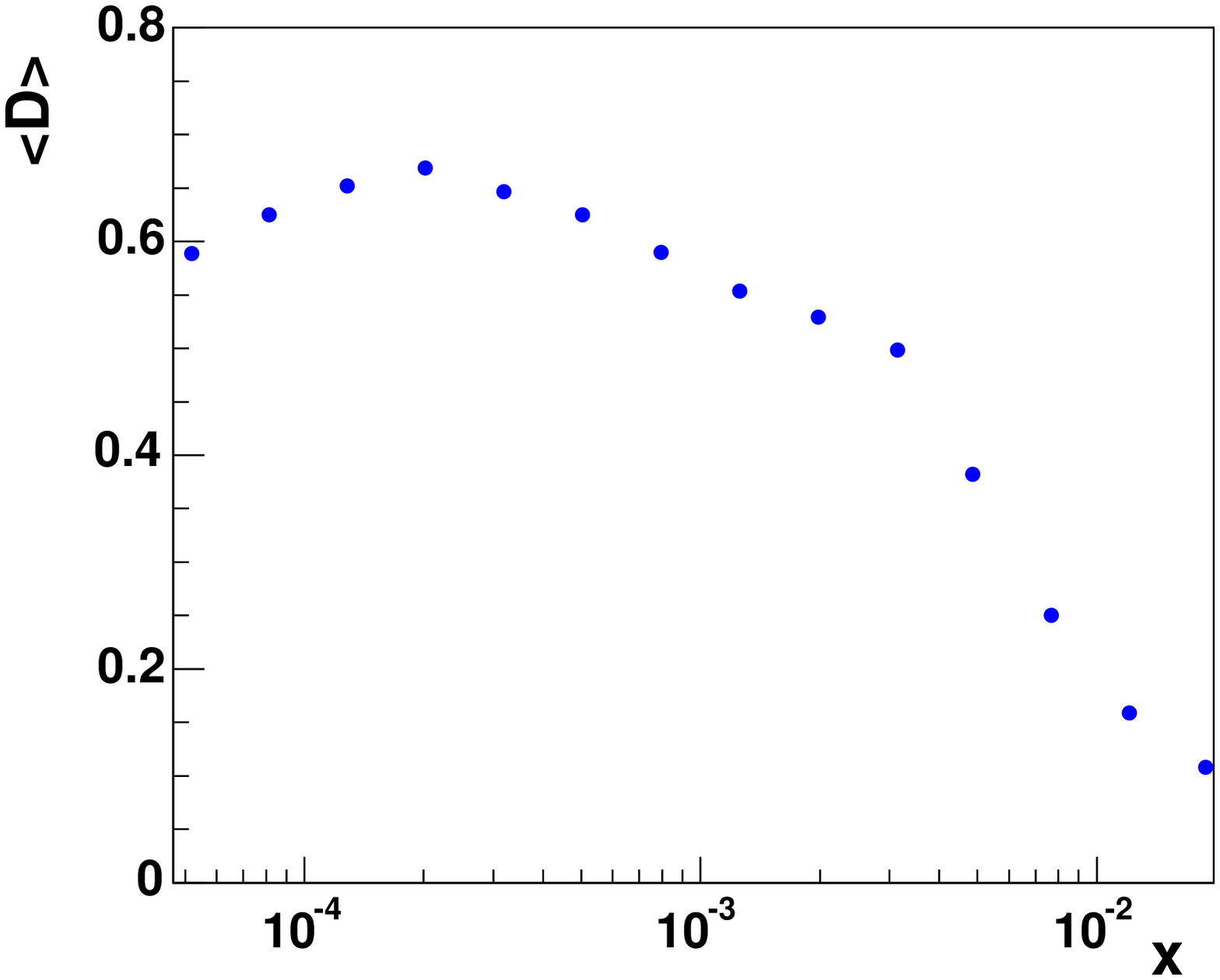}
    \caption{\small Mean depolarisation factor,  $\langle D\rangle$, as
a function of $x$   for the final data sample.}
    \label{fig:depol}
  \end{minipage}
  \hfill
  \begin{minipage}{0.52\linewidth}
    \hskip-1cm
    \includegraphics[width=\textwidth,clip]{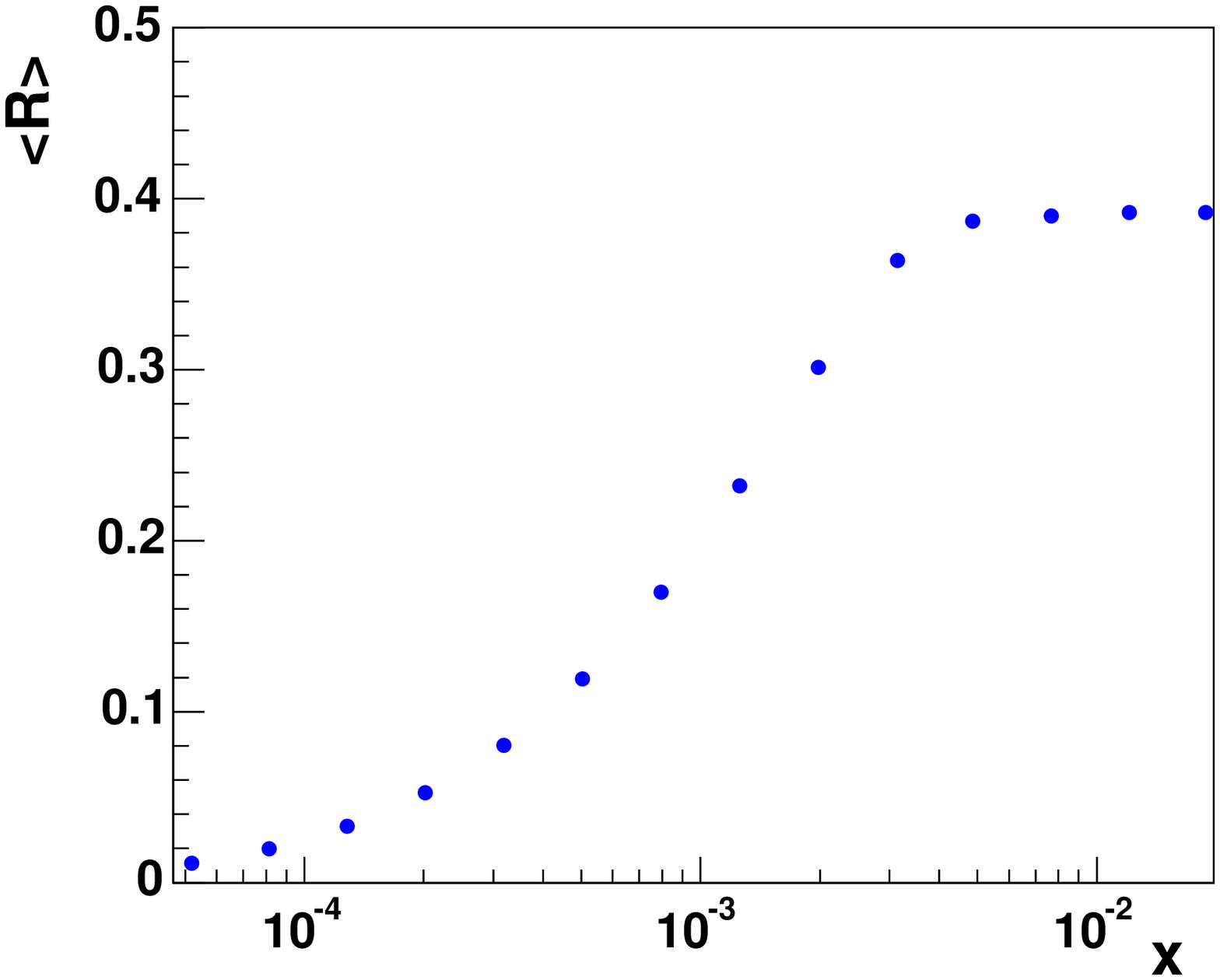}
    \caption{\small Mean values of the ratio $R=\sigma^L/\sigma^T$ 
as a function of $x$ for the final data sample.}
    \label{fig:R}
  \end{minipage}
\end{figure}

In view of the small value of $\eta$ in our kinematic region
the expression for $A^d_1$ in 
Eq.~(\ref{ad}) is reduced to $A^d_1\simeq A^d_\parallel/D$ and the possible
contribution from the neglected term is included in the systematic errors
\cite{ms_phd}.
The virtual-photon depolarisation factor $D$ depends on the 
{ratio of longitudinal and transverse photoabsorption cross-sections, 
$R = \sigma^L/\sigma^T$}. In the present analysis an
updated parametrisation of $R$ taking into account all existing
measurements is  used  \cite{R_1998} together with an extension to
very low values of $Q^2$, cf.\ Appendix. Average values of $D$
and $R$ in bins of $x$ are shown in Figs.~\ref{fig:depol,fig:R},
respectively.

In order to minimize the statistical error of the asymmetry, 
the kinematic factors
$f$, $D$ and {the beam polarisation} $P_B$  are calculated event-by-event and
used to weight events. This approach improves the statistical precision
by approximately 8\% as compared to asymmetry evaluation from events numbers.
In the weight calculations a parametrisation of $P_B$ as a function of 
the beam momentum is used. For $P_T$ an average value is used for the 
data sample taken between two consecutive target spin reversals
\footnote{As $P_T$ varies with time, using it in the
weight would bias the $A_1$ asymmetry.}.
The obtained asymmetry is corrected for spin-dependent
radiative effects  according to Ref.~\cite{polrad} but retaining 
only radiative inelastic tails.

\begin{figure}
\begin{center}
\includegraphics[width=\textwidth,clip]{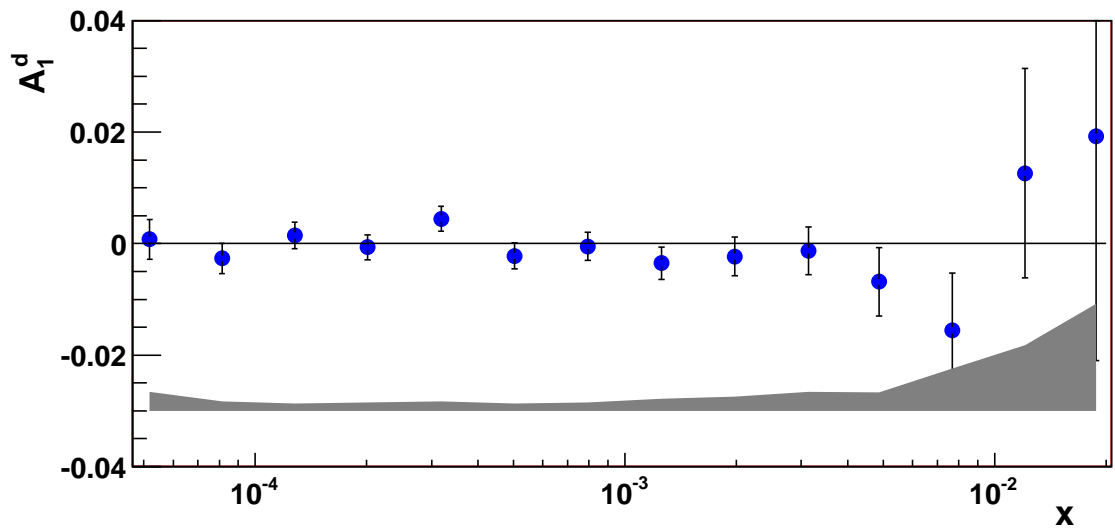}
\end{center}
\caption{\small The asymmetry $A_1^d(x)$ as a function of $x$ at the measured values
of $Q^2$ obtained in this analysis. Errors are statistical; the shaded band 
indicates the size of the systematic ones. }
\label{fig:a1}
\end{figure}

The final values of $A_1^d(x,Q^2)$ 
are listed in Table~\ref{tab:a1_g1} with the
corresponding average values of $x$ and $Q^2$. They are also 
shown as a function of $x$ in Fig.~\ref{fig:a1}. 
These values confirm, with a 
statistical precision increased by more than an order of magnitude, 
the observation made in Ref.~\cite{smc_lowx}  
that the asymmetry is consistent with zero for $x\lapproxeq 0.01$.

The systematic error of $A_1^d$ contains  multiplicative contributions 
resulting from
uncertainties on polarisations $P_B$ and $P_T$, on the dilution factor $f$ 
and on the function $R$ used to calculate the depolarisation factor $D$. Of
these, the largest contribution comes from $D$ due to a poor knowledge
of $R$. When combined in quadrature, these errors amount to 
10--30\% (Table~\ref{tab:sys_error}). 
However the most important contribution to the systematic
error is due to possible false asymmetries which could be generated 
by instabilities in some components of the spectrometer. 
In order to minimize their effect, the values of $A_1^d$ in each interval of
$x$ have been calculated for 97 subsamples, each of them covering a short
period of running time and, therefore, ensuring similar detector operating
conditions. An upper limit of the effect of the time dependent detector 
instabilities has been evaluated by a statistical approach.
Dispersions of the values of $A_1^d$ around their means at each value
of $x$ were compared with their expected values. 
Using the Monte Carlo technique for a statistical limit estimate
\cite{helene}, values  for the false asymmetries were calculated and everywhere 
found to be smaller than the statistical precision. This estimate accounts 
for the time variation effects of the spectrometer components.

Several other searches for false asymmetries were performed. Data from the
two target cells were combined in {different ways} in order to eliminate the
spin-dependent asymmetry. Data obtained with 
opposite signs of cell polarisations 
were compared as they may reveal acceptance effects. 
These searches did not show any significant false asymmetry.

\begin{figure}
\begin{center}
\includegraphics[width=\textwidth,clip]{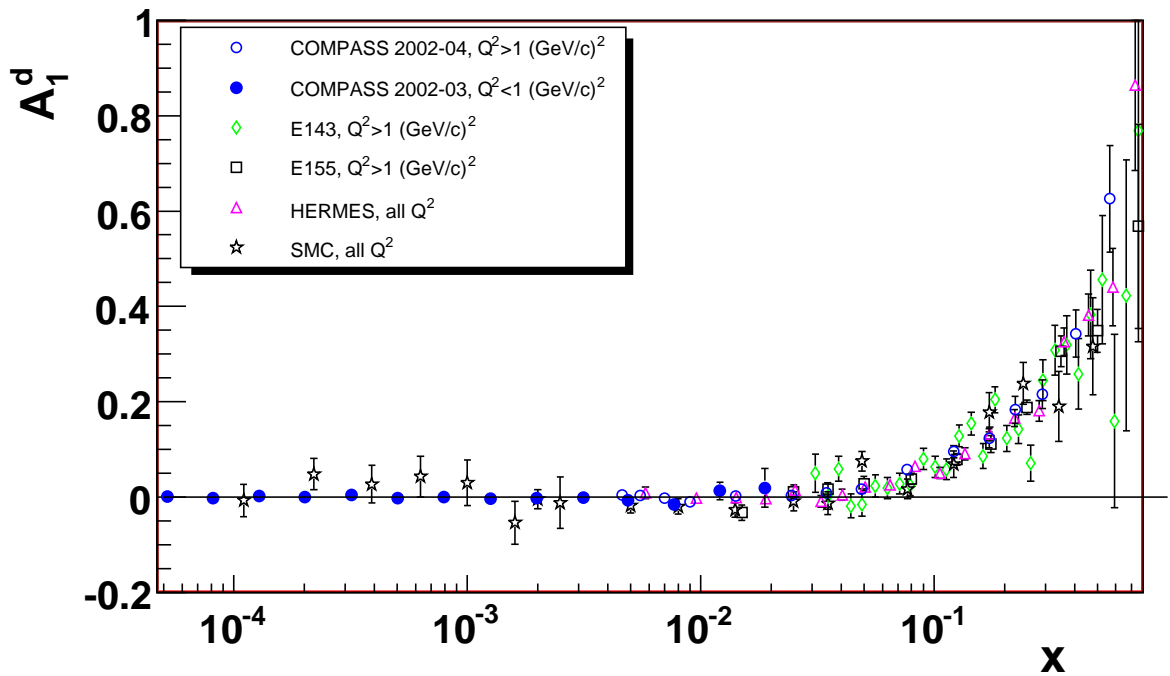}
\end{center}
\caption{\small The asymmetry $A_1^d(x)$ as a function 
of $x$ at the measured values of $Q^2$: the results  for $Q^2 < 1~(\GeV/c)^2$ obtained 
in this analysis are compared with previous results at different 
values of $Q^2$ from COMPASS \cite{compass
_a1_recent}, SMC \cite{smc,smc_lowx}, HERMES \cite{hermes_new}, 
SLAC E143 \cite{e143} and SLAC E155 \cite{e155_d}. 
The E155 data correponding to the same $x$ have been 
averaged over $Q^2$. Errors are statistical. 
}
\label{fig:a1_compar}
\end{figure}

In Fig.~\ref{fig:a1_compar} results of the present analysis 
as a function of $x$ are presented together
with previous measurements by the SMC at 
$0.01 < Q^2 <100~(\GeV/c)^2$ \cite{smc_lowx,smc}.  
The improvement in the statistical precision at low $x$ is striking. 
Other data, mostly from the deep inelastic scattering  region by COMPASS 
\cite{compass_a1_recent}, HERMES \cite{hermes_new}, SLAC E143 \cite{e143}
and SLAC E155 \cite{e155_d}, 
are also presented in Fig.~\ref{fig:a1_compar}.
The values of $A_1^d$, even if  originating from experiments at
different energies, tend to coincide due to the very small $Q^2$ dependence of
$A_1^d$ at fixed $x$.

\begin{figure}
\begin{center}
\includegraphics[width=\textwidth,clip]{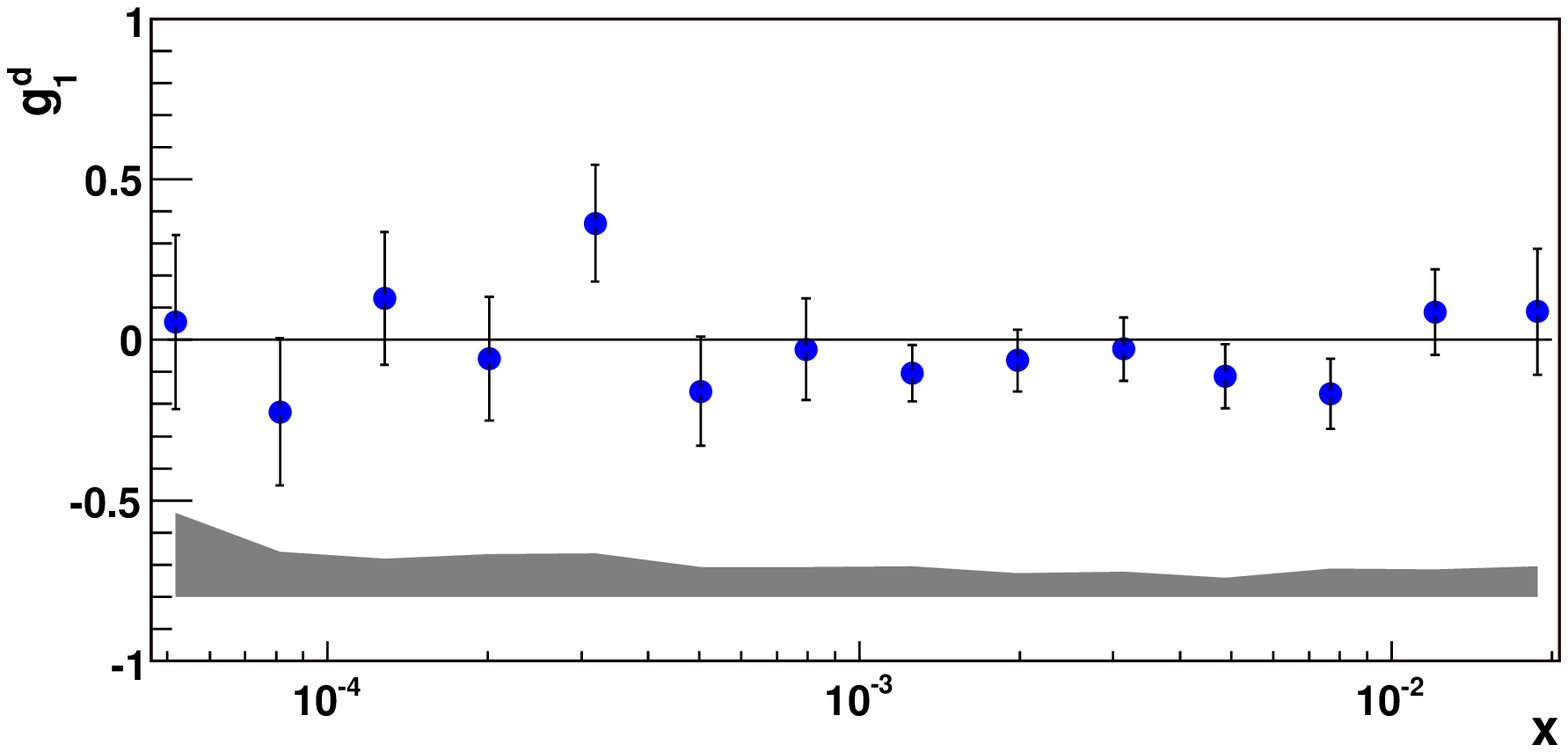}
\end{center}
\caption{\small The spin dependent structure fnction  $g_1^d(x)$ as a function 
of $x$ at the measured values of $Q^2$ obtained in this analysis. Errors are 
statistical; the shaded band indicates the size of the systematic ones. }
\label{fig:g1}
\end{figure}

\begin{figure}
\begin{center}
\includegraphics[width=\textwidth,clip]{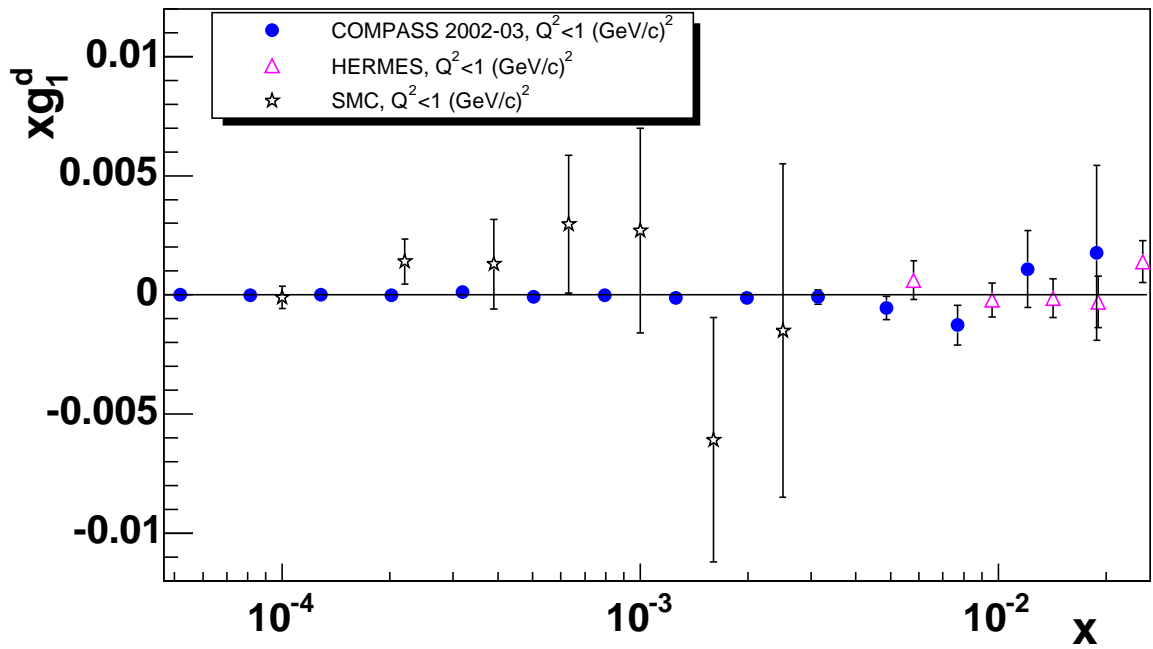}
\end{center}
\caption{\small Same as in Fig.~\ref{fig:a1_compar} but for the 
quantity $xg_1^d$. Only data for $Q^2 <1~(\GeV/c)^2$ are shown.}
\label{fig:g1_compar}
\end{figure}

The spin dependent structure functions are connected to the virtual
photon asymmetries in the following way
\begin{equation}
g_1^d = {F_1^d\over (1 + \gamma^2)}\left (A_1^d + \gamma A_2^d\right ),
\hskip2cm 
g_2^d = {F_1^d\over (1 + \gamma^2)}\left (-A_1^d + {1\over\gamma} A_2^d\right ).
\end{equation}
These formulae are exact; possible contributions from the structure functions
$b_{1-4}$ cancel out.
Neglecting $A_2^d$ and making
the usual replacement $(1 + \gamma^2)F_2/(2xF_1) = 1 + R$,
as in the spin-1/2 case and valid if $b_{1-4} = 0$,
the longitudinal spin structure function $g_1^d$ is obtained as
\begin{equation}
g_1^d = \frac{F_2^d}{2~ x~(1 + R)} A_1^d.
\end{equation} 
The values of $g_1^d$ are listed in the last column of Table~\ref{tab:a1_g1}
and shown in Fig.~\ref{fig:g1}.
They have been obtained with the $F_2^d$ parametrisation of 
Refs.~\cite{smc,jkbb_f2}, cf.\ Appendix,
and with the parametrisation of $R$ used in the depolarisation factor. 
The systematic errors on $g_1^d$ are obtained in the same way as for $A_1^d$,
with an  additional contribution from the uncertainty on $F_2^d$.
Moreover the error of the depolarisation factor was modified. Instead 
of $\delta D$,
the error of the quantity $D(1 + R)$, $\delta [D(1 + R)]$ was considered.
The values of $xg_1^d(x)$ obtained in this analysis and, for comparison, 
the SMC \cite{smc_lowx} and HERMES \cite{hermes_new} results at 
$Q^2 < 1~(\GeV/c)^2$ are shown in Fig.~\ref{fig:g1_compar}.

The low $x$ data in the kinematic region where $W^2$ is high 
and $W^2\gg Q^2$, should in principle allow testing the Regge behaviour
of $g_1$ through its $x$ dependence. These conditions are fulfilled 
by our measurements and thus
 a fit of Eq.~(\ref{regge}) to the $g_1$ data from the $Q^2$ range of
$0.0025\ -\ 0.25~(\GeV/c)^2$ in six subintervals of 
$Q^2 \approx \mbox{const}$ was performed. The results of the fit were inconclusive. 
No information on the singlet intercept, $\alpha_s(0)$, could be extracted. 
Thus our data do not provide a test of 
the Regge behaviour of $g_1$ without additional assumptions about its
$Q^2$ dependence. This is due to a limited $x$ interval for 
any given value of $Q^2$ combined with small measured values of $g_1^d$. 
However, these data can be compared with models 
predicting both the $x$ and $Q^2$ dependence of $g_1$ 
at low values of $x$ and $Q^2$ \cite{greco,nonpert}. 
A relevant phenomenological analysis is in progress.

In summary, we have measured the deuteron spin asymmetry $A_1^d$ 
and its longitudinal spin-dependent structure function $g_1^d$ 
for $Q^2<1~(\GeV/c)^2$ over the range 
$4\cdot 10^{-5} < x < 2.5\cdot10^{-2}$ and with a statistical precision
more than tenfold better than previous experiments. 
The $A_1^d$ and $g_1^d$ values are compatible with zero for 
$ x\lapproxeq 0.01$.

\section*{Acknowledgements} 

\vskip5mm

We gratefully acknowledge the support of the CERN management and staff and  
the skill and effort of the technicians and engineers of our collaborating 
institutes. 
Special thanks are due to V.~Anosov, J.-M. Demolis and V.~Pesaro for 
their technical support
during the installation and the running of this experiment. 
This work was made possible by the financial support of our funding agencies.

\section*{Appendix}

\vskip5mm

Knowledge of $F_2(x,Q^2)$ and $R(x,Q^2)$ is needed in computations 
of the dilution factor, 
the radiative corrections, the depolarisation factor and the
spin dependent structure function $g_1(x,Q^2)$. It is not 
sufficient to know these functions only in the kinematic range of the
analysis since
radiative corrections require their knowledge at $x > x_{meas}$  and
all values of $Q^2$ including $Q^2 = 0$,
due to radiative ``tails''.
Asymptotic behaviours of $F_2$ and $R$ in the photoproduction limit, 
$Q^2 \rightarrow 0$, are: $F_2\sim Q^2$ and $R\sim Q^2$ (for fixed, 
arbitrary $\nu$).
These kinematic constraints eliminate potential kinema\-ti\-cal
singularities at $Q^2 = 0$ of the hadronic tensor defining the
virtual Compton scattering amplitude.

In the analysis, a new SLAC parametrisation of $R$, $R_{1998}$ \cite{R_1998}, 
and $F_2$ parametrisation of Ref.~\cite{smc}  is employed.
The former, valid for $Q^2 >0.5~(\GeV/c)^2$, is  extended to lower
values of $Q^2$, including the  $R \sim Q^2$ behaviour at  $Q^2 = 0$, as:
\begin{equation}
R(Q^2 < 0.5~(\GeV/c)^2, x) = R_{1998}(0.5~(\GeV/c)^2, x) 
\cdot \beta(1 - \exp{(-Q^2/\alpha)})
\end{equation}
with $\alpha = 0.2712$ and $\beta = 1/(1 - \exp(0.5/\alpha)) = 1.1880$.
At $Q^2 = 0.5~(\GeV/c)^2$ the function and its first derivative are
continuous in the whole $x$ range of our measurements.  
The error on $R$, $\delta R$, above $Q^2 = 0.5~(\GeV/c)^2$ was taken from
Ref.~\cite{R_1998} and below $Q^2 = 0.5~(\GeV/c)^2$ was set to 0.2.
For this value of $\delta R$ and for the simplest assumption
about $R$ at $Q^{2}<0.5~(\GeV/c)^2$ and any $x$ ({\it e.g.} $R=0.2$) there is
an approximate agreement (within 1$\sigma$)  with both the value $R = 0$ 
at the photoproduction limit
and with measurements at higher $Q^{2}$ from HERA, where $R \approx 0.4$
 \cite{R_HERA}.

The $F_2$ of Ref.~\cite{smc} is valid for $Q^2 > 0.2~(\GeV/c)^2$ and $x > 0.0009$. 
At lower values of $Q^2$ and $x$ we used the model of Ref.~\cite{jkbb_f2}
valid down to $Q^2 = 0$ and $x = 10^{-5}$ and based on a concept
of generalised vector meson dominance. Two other
$F_2$ parametrisations, albeit for the proton \cite{f2_saturation,ALLM97}, 
were also tried together with Ref.~\cite{jkbb_f2},
to estimate the $F_2$ uncertainty, $\delta F_2$. The former of these 
parametrisations is based on the parton saturation model with recent 
modi\-fi\-ca\-tions including the QCD evolution and the latter
is a Regge motivated fit to all the world data of $F_2^p$, extended into
the large $Q^2$ in a way compatible with QCD expectations. They are 
valid in a range similar to that of Ref.~\cite{jkbb_f2}. The $\delta F_2$
 uncertainty was taken as the largest difference
between the values of the employed $F_2$ and other parametrisations.

\begin{table}
\begin{center}
{\footnotesize
\begin{tabular}{|l@{--}l|c|c|c|r|r|}
\hline \hline
\multicolumn {2}{|c|}{}&&&&& \\
\multicolumn {2}{|c|}{$x$ range} & $\langle x \rangle$ & $\langle Q^2 \rangle $  & $\langle y\rangle $  & \multicolumn{1}{c|}{$A_1^d$} & \multicolumn{1}{c|}{$g_1^d$} \\
\multicolumn {2}{|c|}{}   &  &   [(GeV$/c$)$^2$]&  &  &  \\
\hline \hline
 0.000063&~~0.00004 & 0.000052 & 0.0068 & 0.44& 0.0008 $\pm$ 0.0036 $\pm$ 0.0034 & $ 0.06 \pm 0.27 \pm 0.26$\\
 0.00004&~~0.0001 & 0.000081 & 0.012~~ & 0.49&$-0.0027 \pm 0.0027 \pm 0.0017$ & $-0.22 \pm 0.23 \pm 0.14$ \\
 0.0001&~~0.00016 & 0.00013~~ & 0.021~~ & 0.53 & 0.0015 $\pm$ 0.0023 $\pm$  0.0014 & 0.13 $\pm$ 0.21 $\pm$  0.12\\
 0.00016&~~0.00025 & 0.00020~~ & 0.034~~ & 0.56 & --0.0007 $\pm$ 0.0022 $\pm$  0.0015& --0.06 $\pm$ 0.19 $\pm$  0.13 \\
 0.00025&~~0.0004 & 0.00032~~ & 0.054~~ & 0.56 & 0.0045 $\pm$ 0.0022 $\pm$  0.0017& 0.36 $\pm$ 0.18 $\pm$  0.14\\
 0.0004&~~0.00063 & 0.00050~~ & 0.085~~ & 0.56 & --0.0022 $\pm$ 0.0023 $\pm$  0.0013& --0.16 $\pm$ 0.17 $\pm$  0.09\\
 0.00063&~~0.001   & 0.00079~~ & 0.13~~~~ & 0.55 & --0.0005 $\pm$ 0.0025 $\pm$  0.0015 & --0.03 $\pm$ 0.16 $\pm$  0.09\\
 0.001&~~0.0016  & 0.0013~~~~ & 0.20~~~~ & 0.54 & --0.0035 $\pm$ 0.0029 $\pm$  0.0022 & --0.11 $\pm$ 0.09 $\pm$  0.09\\
 0.0016&~~0.0025  & 0.0020~~~~ & 0.32~~~~ & 0.54 & --0.0023 $\pm$ 0.0035 $\pm$  0.0025 & --0.07 $\pm$ 0.10 $\pm$  0.07\\
 0.0025&~~0.004   & 0.0031~~~~ & 0.50~~~~ & 0.53 & --0.0013 $\pm$ 0.0043 $\pm$  0.0034 & --0.03 $\pm$ 0.10 $\pm$  0.08\\
 0.004&~~0.0063  & 0.0049~~~~ & 0.63~~~~ & 0.43 & --0.0069 $\pm$ 0.0061 $\pm$  0.0033& --0.11 $\pm$ 0.10 $\pm$  0.06\\
 0.0063&~~0.01    & 0.0077~~~~ & 0.68~~~~ & 0.30 &--0.016~~ $\pm$ 0.010~~ $\pm$  0.008~~~& --0.17 $\pm$ 0.11 $\pm$  0.09\\
 0.01&~~0.0158  & 0.012~~~~~~ & 0.74~~~~ & 0.20 &0.013~~ $\pm$ 0.019~~ $\pm$  0.012~~~& 0.09 $\pm$ 0.13 $\pm$  0.09\\
 0.0158&~~0.025   & 0.019~~~~~~ & 0.82~~~~ & 0.14 & 0.019~~ $\pm$ 0.040~~ $\pm$  0.019~~ & 0.09 $\pm$ 0.20 $\pm$  0.09\\
\hline \hline
\end{tabular}
}
\caption{\small
  Values of $A_1^d$ and $g_1^d$ with their statistical and systematic errors
  as a function of $x$ with the corresponding average values of $x$, $Q^2$ and $y$.
  The maximum $Q^2$ cut is 1 (GeV$/c$)$^2$. Bins in $x$ are of equal width 
in log$_{10}x$.}
\label{tab:a1_g1}
\end{center}
\end{table}

\begin{table}
\begin{center}

\begin{tabular}{|l|l|c|c|}
\hline
\hline
              & Beam polarisation     & $\delta P_B/P_B$ & 4\% \\
\cline{2-4}
Multiplicative & Target polarisation   & $\delta P_T/P_T$ & 5\% \\
\cline{2-4}
variables      & Depolarisation factor & $\delta D(R)/D(R)$ & 4 -- 30 \% \\
\cline{2-4}
error& Dilution factor & $\delta f/f$    & 7 \% \\
\hline
\hline
Additive       & Transverse asymmetry  & ${\eta}\cdot\delta A_2$ & $<
0.1\cdot\delta A_1^{stat}$ \\
\cline{2-4}
variables      & Radiative corrections& $\delta A_1^{RC}$ &
$< 0.03 \cdot\delta A_1^{stat}$\\
\cline{2-4}
error & False asymmetry   & $A_{false}$ & $<\delta A_1^{stat}$ \\
\hline
\hline
\end{tabular}
\end{center}
\caption{\small Decomposition of the systematic error of $A_1$
into multiplicative and additive variables contributions.}
\label{tab:sys_error}
\end{table}

\end{document}